\documentclass[fleqn,usenatbib]{mnras}

\usepackage{newtxtext,newtxmath}

\usepackage{graphicx}

\usepackage[T1]{fontenc}

\DeclareRobustCommand{\VAN}[3]{#2}
\let\VANthebibliography\thebibliography
\def\thebibliography{\DeclareRobustCommand{\VAN}[3]{##3}\VANthebibliography}

\usepackage{graphicx}	
\usepackage{amsmath}	

\title[Rotating Black Hole Lensing]{Observational Signatures of Kerr Black Hole Lensing}

\author[T. P. Kling et al.]{
Thomas P. Kling ,$^{1}$\thanks{E-mail: tkling@bridgew.edu}
Lia Troy,$^{2}$
Shannon Harding$^{3}$
and Timothy O'Leary$^{1}$
\\
$^{1}$Dept. of Physics, Photonics and Optical Engineering, Bridgewater State University,
Bridgewater, MA 02325 USA\\
$^{2}$Dept. of Mathematics, University of Chicago, Chicago, IL 60637, USA  \\
$^{3}$Dept. of Physics and Applied Physics, University of Massachusetts-Lowell, Lowell, MA, 01814, USA
}

\date{Pubished: October 2026, MNRAS}

\pubyear{\the\year{}}

\begin{document}
\label{firstpage}
\pagerange{\pageref{firstpage}--\pageref{lastpage}}
\maketitle

\begin{abstract}
Gravitational lensing by rotating black holes is modeled, and the null geodesic equations in the strong field regime are integrated numerically. Observational signatures of the black hole spin parameter are identified by comparing image locations and shapes of extended sources at different $a$ values. For the black hole at the center of our galaxy, angular resolutions of the order of hundreds of micro arc seconds are needed to distinguish between the true source position and gravitationally lensed image position. Angular resolution below one micro arc second is needed to determine the spin parameter of the galactic black hole using gravitational lensing.
\end{abstract}




\section{Introduction} \label{intro:sec}

Recent advances in observational techniques have led to the first accurate images of the central region of our galaxy, revealing the shadow of the black hole (\cite{Akiyama_2019}), and highly accurate maps of the motion of stars very close to the black hole using the GRAVITY interferometer at the European Southern Observatory are testing predictions of general relativity such as the Schwarzschild precession (\cite{refId0}). These advances, and planned future improvements, offer the possibility of examining key features of Einstein's Theory of General Relativity beyond the proof that at the center of our galaxy is a supermassive black hole.

Images we see from sources of light near the center of our galaxy will be distorted by the strong gravitational fields present. Typical gravitational lensing studies assume weak gravitational fields and large distances between the source of light, lens, and observer. While we are very far from the center of our galaxy, objects orbiting Sagittarius A* (Sgr A*) are relatively close and may allow future missions to probe gravitational lensing by these strong fields. Current studies of the motion of stars near Sgr A* do not take into account gravitational lensing effects of the black hole.

Black hole gravitational lensing, where one develops lens equations and observational consequences of general relativity, is based on the geodesic equations and geodesic deviation equations. The formalism was initially developed by \cite{virbhadra} and \cite{ehlersNewman}, with details worked out for Schwarzschild black holes by other authors (\cite{FKN}). Broadly, the approach is referred to as \textit{exact} gravitational lensing, as opposed to utilizing the more common thin lens approximations, and it is based on treating lensing as an examination of the past light cone of the observer (\cite{Perlick}). 

In terms of Kerr (rotating) black holes, an early analysis of Kerr-type black holes caustics was performed by \cite{Rauch}. \cite{bozza01} and \cite{sereno03} independently worked to develop Kerr black hole lensing. \cite{aazami} introduced a set of lens equations for Kerr metrics, and photon trajectories in extreme Kerr black holes ($a>m$) were considered by \cite{porfy}. \cite{beckwith} worked through many observational consequences of black hole lensing on the accretion disk. \cite{Bozza05} continues this work, examining the structure of caustics and their relation to images. Several other recent papers examine the images of sources near rotating black holes, deriving image location (\cite{Hsieh}), examining the black hole shadow (\cite{kuang}), and determining the appearance of a star in orbit of a high spin parameter black hole (\cite{doku}). The formalism of integrating light rays near Kerr black holes using the Carter approach was more recently re-examined in \cite{Gralla2} and \cite{Gralla1}.

The purpose of this analysis is to examine potentially observable signatures of gravitational lensing of Sgr A* assuming the system is a rotating (Kerr) black hole. We are particularly interested in whether observations of gravitational lensing can help determine the spin parameter, $a$, of the Kerr black hole. As such, we provide visualization of gravitational lensing effects at both $a=0$ and high spin ($a=0.9$), as well as tables of potential observations at a different values of $a$.

This paper is organized as follows. First, in Section~\ref{kerr:sec}, we provide an outline of the Kerr metric and explain the broad approach taken to achieve gravitationally lensed images. (Further details are provided in Appendices.) Section~\ref{pointsandflyby:sec} shows the broad effect of lensing on point sources behind a rotating black hole and examines the appearance of an object that moves across the sky in a linear fashion. In Sec.~\ref{strong_lesnsing_sec}, we discuss how extended objects are lensed into arcs, comparing those arcs for different values of the spin parameter $a$. In this section, we see that the spin of the black hole shifts the locations of the two lensed arcs so that the line connecting the arcs does not pass through the black hole when $a\ne 0$. Finally, in Sec.~\ref{orbit:sec}, we examine the changes due to lensing of the image location relative to the source location for an object orbiting the black hole, and provide estimates of this shift in position for a star in close orbit of Sgr A*.


\section{Kerr Metric} \label{kerr:sec}

The most common expression of the spacetime metric for a rotating black hole is given using the Boyer and Lindquist coordinates (\cite{hawkingellis}). The metric in these coordinates is given by

\begin{eqnarray}  ds^2 &=& \rho^2 \left( \frac{dr^2}{\Delta} + d\theta^2 \right) + (r^2 + a^2) \sin^2 \theta d\phi^2 - dt^2 \nonumber \\ & ~ & ~ + \frac{2mr}{\rho^2} (a\sin^2\theta d\phi - dt)^2 \label{kerr1}, \end{eqnarray}

\noindent where $\rho$ and $\Delta$ are defined in terms of the mass $m$ and spin parameter $a$ by

\begin{eqnarray} \rho^2(r, \theta) &= & r^2 + a^2 \cos^2 \theta \\ \Delta & = & r^2 - 2mr + a^2. \end{eqnarray}

\noindent The geodesic equations for light rays of the Kerr metric possesses first integrals of the motion in these coordinates. Beginning with Carter (1968), most approaches to gravitational lensing in the Kerr metric make use of these first integrals to write out elliptic integrals, which when solved numerically provide light ray trajectories (\cite{carter}). 

The approach taken in this paper is different, and specialized to integrating a large number of light rays. We choose to numerically integrate the geodesic equations as ordinary differential equations and start by changing to the Kerr coordinates, introducing the variables $u_-$ and $\phi_-$ through

\begin{eqnarray}  du_- &=& dt - \frac{(r^2 + a^2)}{\Delta} dr \label{uminusdef} \\ d\phi_- &= & d\phi - \frac{a}{\Delta} dr. \label{phiminusdef}\end{eqnarray}

\noindent In the $(u_-, r, \theta, \phi_-)$ coordinates, the line element of the metric $g_{ab}$ is 

\begin{eqnarray} ds^2  & = & \rho^2 d\theta^2   +  2 a \sin^2\theta dr d\phi_- - 2 dr du_- + \rho^{-2}[(r^2+a^2)^2 \nonumber \\ & ~ & ~ - \Delta a^2 \sin^2 \theta] \sin^2\theta d\phi_-^2  -4a\rho^{-2} m r \sin^2\theta d\phi_- du_- 
\nonumber \\ & ~ & ~- (1-2mr\rho^{-2}) du_-^2. \label{kerr2} \end{eqnarray}

For our numerical approach, we introduce a Lagrangian

\begin{equation}
    \mathcal{L} = \frac{1}{2} g_{ab} \dot x^a \dot x ^b , \label{lag_gen}
\end{equation}

\noindent and in \cite{klingKerr1} a lengthy calculation showed that the resulting four second-order Euler-Lagrange equations can be decoupled into eight first-order ordinary differential equations. The final equations are provided in Appendix A along with a discussion of the appropriate initial conditions. This previous work demonstrated that the resulting differential equations could be integrated using highly-accurate adaptive step-size techniques very close to the black hole event horizon. The strategy we use for identifying specific light rays connecting sources and the observer is described in Appendix B. Validation and comparison with the Carter approach is discussed in Appendix C.

Throughout this paper, we will consider two sets of lensing scenarios. In the first, we use a generic lensing case with distance scales chosen primarily to assist in visualizing the lensing effects. In the visualization scenario, we use dimensionless units where $Gm/c^2 = 1$, and either $a=0$ (no rotation) or $a=0.9$ (high rotation). To examine general lensing, we set the observer as $1000$ dimensionless units in front of the black hole and source object $1000$ dimensionless units behind the black hole. At these distances and scales, the source and observer are very nearly in the asymptotic region of the spacetime, but we will still be able to visualize the impact of lensing. When we consider an object orbiting the black hole, we use a observer $1000$ dimensionless units away from the black hole and allow the orbiting source to maintain a distance of $50$ dimensionless units. Figures presented when considering visualization will have axes whose units are in terms of the distance scale $Gm/c^2=1$.

As a second set of lensing scenarios, we present results in realistic lensing scales. Here we use approximate mass of Sgr A* and set the observer distance as our distance from the center of our galaxy ($8.3$~kpc). We examine sources at either $120$ AU or $360$ AU from Sgr A*, noting that the distance of closest approach of the star S2 is $120$ AU, even though this star has a highly elliptical orbit. In this second scenario, the distance of an observer on Earth to the black hole is $3.87\times10^{10}$ times the mass of the black hole, and $120$ AU is $2820$ times the mass of the black hole. Figures and tables presented in this realistic lensing scale make use of angular variables where the angular distances are converted from physical distances by dividing by the distance to Sgr A*.


\section{Kerr Lensing of Point Sources and Flybys} \label{pointsandflyby:sec}

\begin{figure*}
\begin{center}
\scalebox{0.40}{\includegraphics{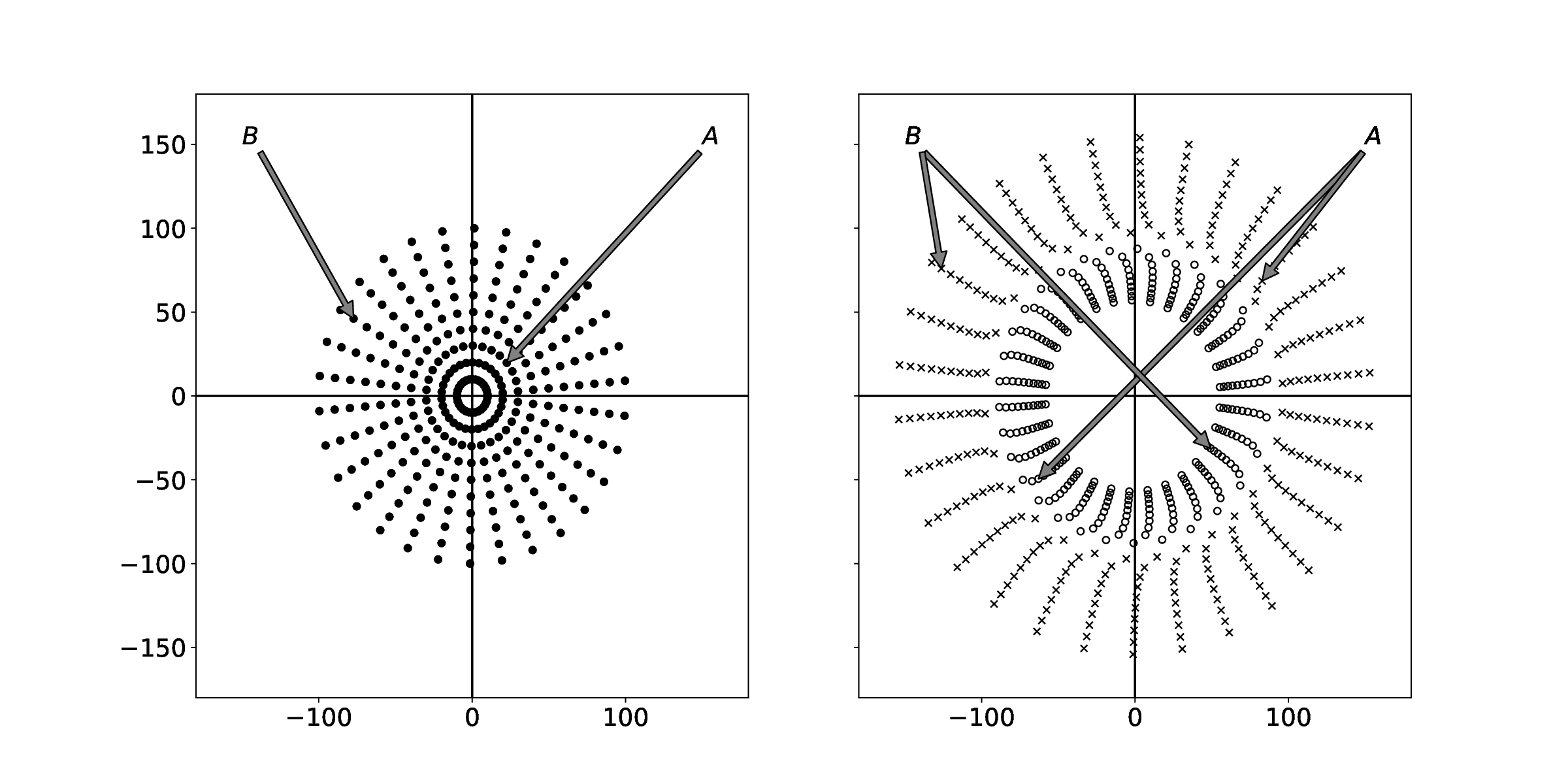}} \caption{\label{SkySource:fig}
Source positions (left pane) and their gravitationally lensed image positions (right pane) for sources at a uniform distance behind a Kerr lens with $a=0.9$. Source positions $A$ and $B$ and their images are labeled. Images marked with a cross are on the same side as the source, and images marked with an open circle are on the opposite side of the black hole.} \end{center}\end{figure*}

As a test that the numerical integration is yielding good results, we created a sky of sources in a single source plane, oriented along circles concentric with the optical axis connecting the observer and center of the black hole. Figure~\ref{SkySource:fig} shows source locations (left pane) with two corresponding image locations (right pane) in the visualization scale. We see that while the source positions lie on straight radial arms, the image locations show evidence of the rotation (with $a=0.9$). Generally, a primary image (same side as the source) is located slightly further away from the optical axis (the small crosses in Fig.~\ref{SkySource:fig}), and the arms on which the primary images lie bend. Secondary images (open circles) have passed by the black hole and are located on the opposite side of the black hole away from the source. The arms along which the secondary sources lie bends in the opposite direction as that of the primary image. The further the source is located from the optical axis, the closer to the black hole its secondary image is seen. Hence, the secondary images of sources further away from the black hole optical axis experience stronger gravity as they pass closer to the event horizon. (See the sources and images labeled $A$ and $B$.)

\begin{figure}
\begin{center}
\scalebox{0.4}{\includegraphics{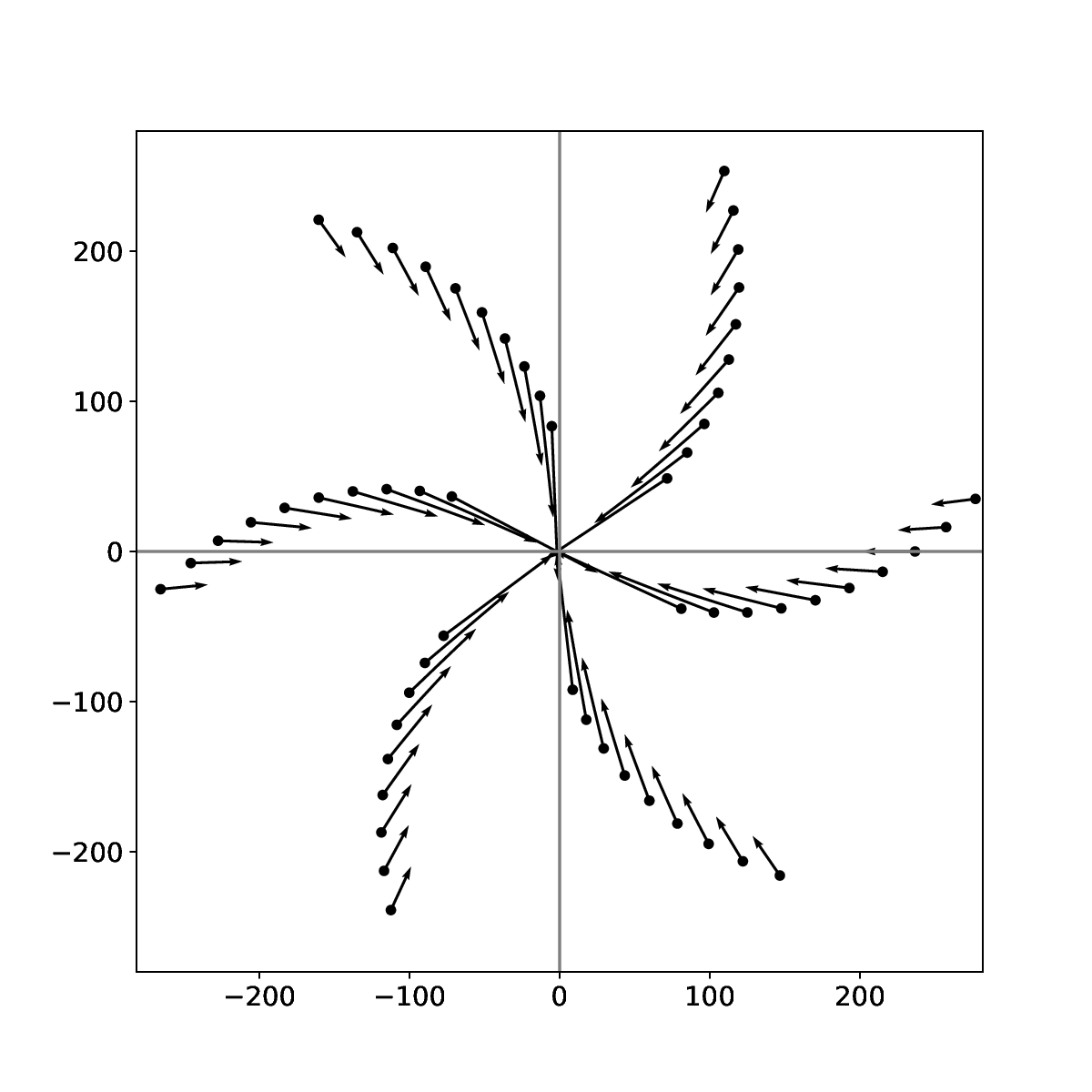}} \caption{\label{viewShifts:fig}
The shift between the true location of the source of light (at the tip of the arrow) and the viewed location (dot) for $a=0.9$ in the visualization scale. Arrows do not all point radially due to the black hole spin.} \end{center}\end{figure}

Figure~\ref{viewShifts:fig} shows the shift in primary image location for a series of source locations in the visualization scale. In this figure, the observed image location is the dot, and the tip of the arrow is the location of the source point. Source points nearly directly behind the black hole are shifted significantly further away from the optical axis. The rotation of the black hole is apparent in the fact that the arrows present do not all point radially.

\begin{figure}
\begin{center}
\scalebox{0.40}{\includegraphics{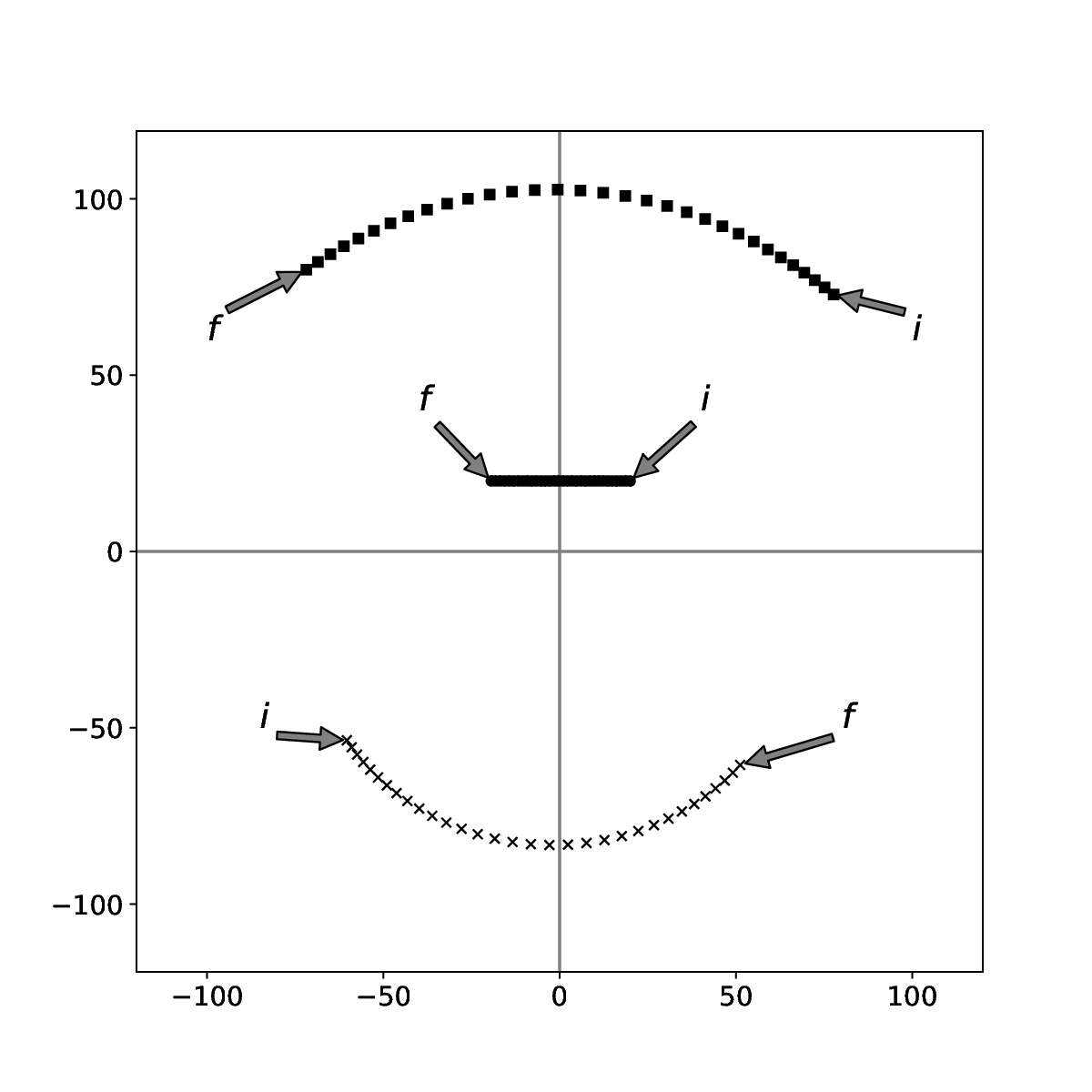}} \caption{\label{pathsz60:fig}
Gravitationally lensed images of an object moving along a horizontal line from the position marked $i$ to the position marked $f$ across the sky. Square boxes locate the primary image on the same side of the black hole as the source, while x marks images that have passed around the black hole to be viewed on the opposite side as the original source. } \end{center}\end{figure}

Finally, we consider a flyby scenario, where a source object is thought to move across the sky. Figure~\ref{pathsz60:fig} shows the primary and secondary images of a source that moves from $y = -20$ to $y = +20$ (horizontal axis), at $z=+20$ (vertical axis), across the sky, with the observer  $1000$ dimensionless units in front of the lens and the source also $1000$ dimensionless units behind the lens. The $i$ and $f$ labels indicate an initial and final location of the source on the flyby along with the primary images (squares) and secondary images (crosses). The trace of the image path is not perfectly symmetric with the vertical axis due to the rotation of the black hole.

\section{Gravitational Lensing Arcs} \label{strong_lesnsing_sec}

The rotation of black hole breaks the axial symmetry present in the case of a non-rotating black hole, changing multiple aspects of the appearance of strong lensing images. In this section we discuss how the presence of rotation impacts the position of extended images and their shapes. We present figures using the visualization scale set (with $a=0.9$ and $a=0.0$ and the observer and source $1000$ dimensionless units away from the black hole) and conclude the section with a table of potentially observable features for the black hole at the center of our galaxy.

\begin{figure}
\begin{center}
\scalebox{0.4}{\includegraphics{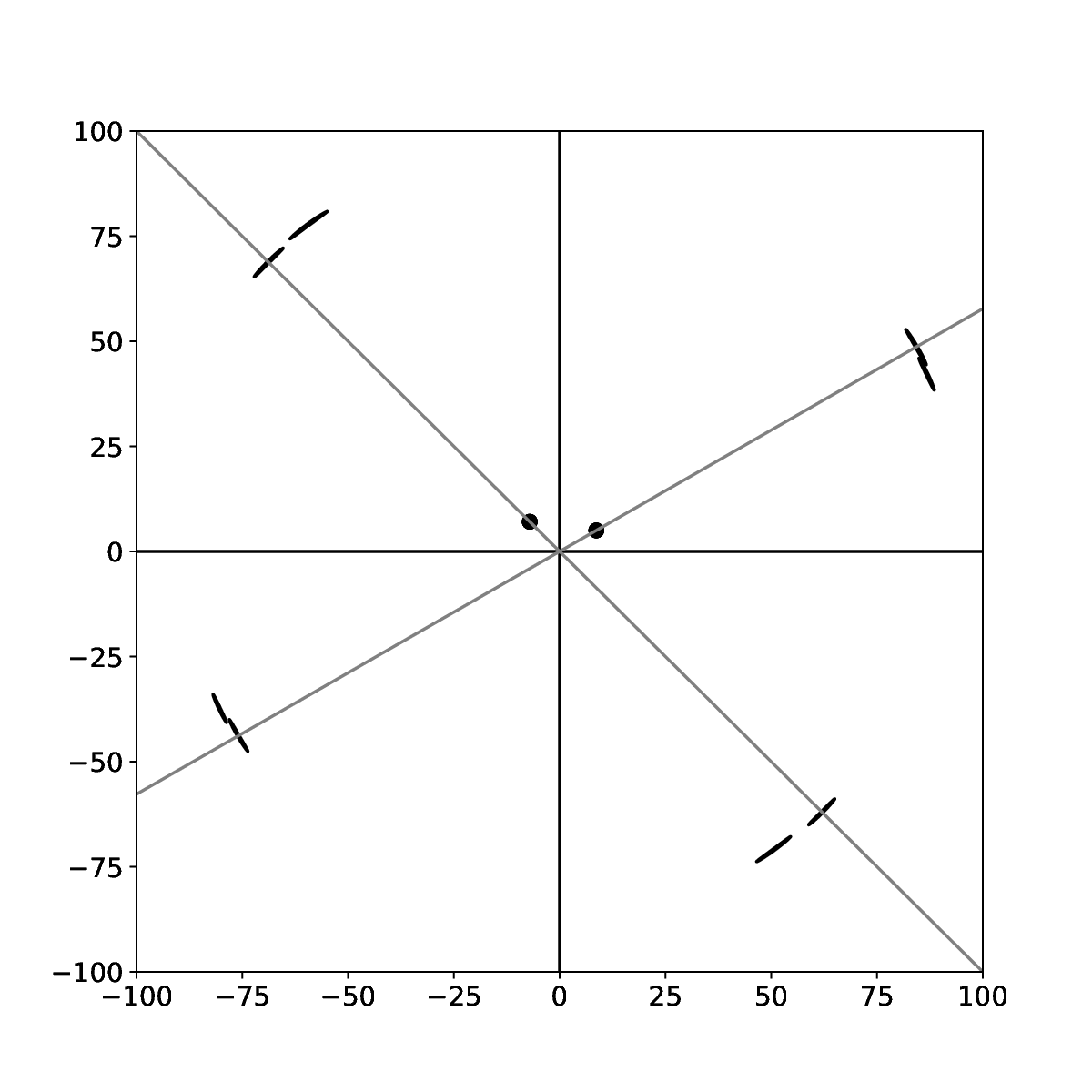}} \caption{\label{twoarcs_new:fig}
Extended images from gravitational lensing created by a black hole at the visualization scale. Sources (shown as solid black circles) are located at a scaled distance of 10 units from the central axis and angles of $\pi/6$ and $3\pi/4$ from the horizontal axis. When $a=0$, the extended lensed images are centered on line passing through the source and central axis. When $a=0.9$, image arcs do not fall along the line of symmetry.} \end{center}\end{figure}

Figure~\ref{twoarcs_new:fig} shows the images of two circular sources as extended arcs in the visualization distance scale. Two sources are shown as small black circles. Both are $10$ dimensionless units away from the central axis, with one rotated $\pi/6$ and other $3\pi/4$ radians counterclockwise from the horizontal. The solid lines pass through the origin where the black hole is centered and the center of the source, and these lines pass through the images for a lens with $a=0$. However, we see that these lines do not pass through the centers of the image arcs when $a=0.9$.

\begin{figure}
\begin{center}
\scalebox{0.4}{\includegraphics{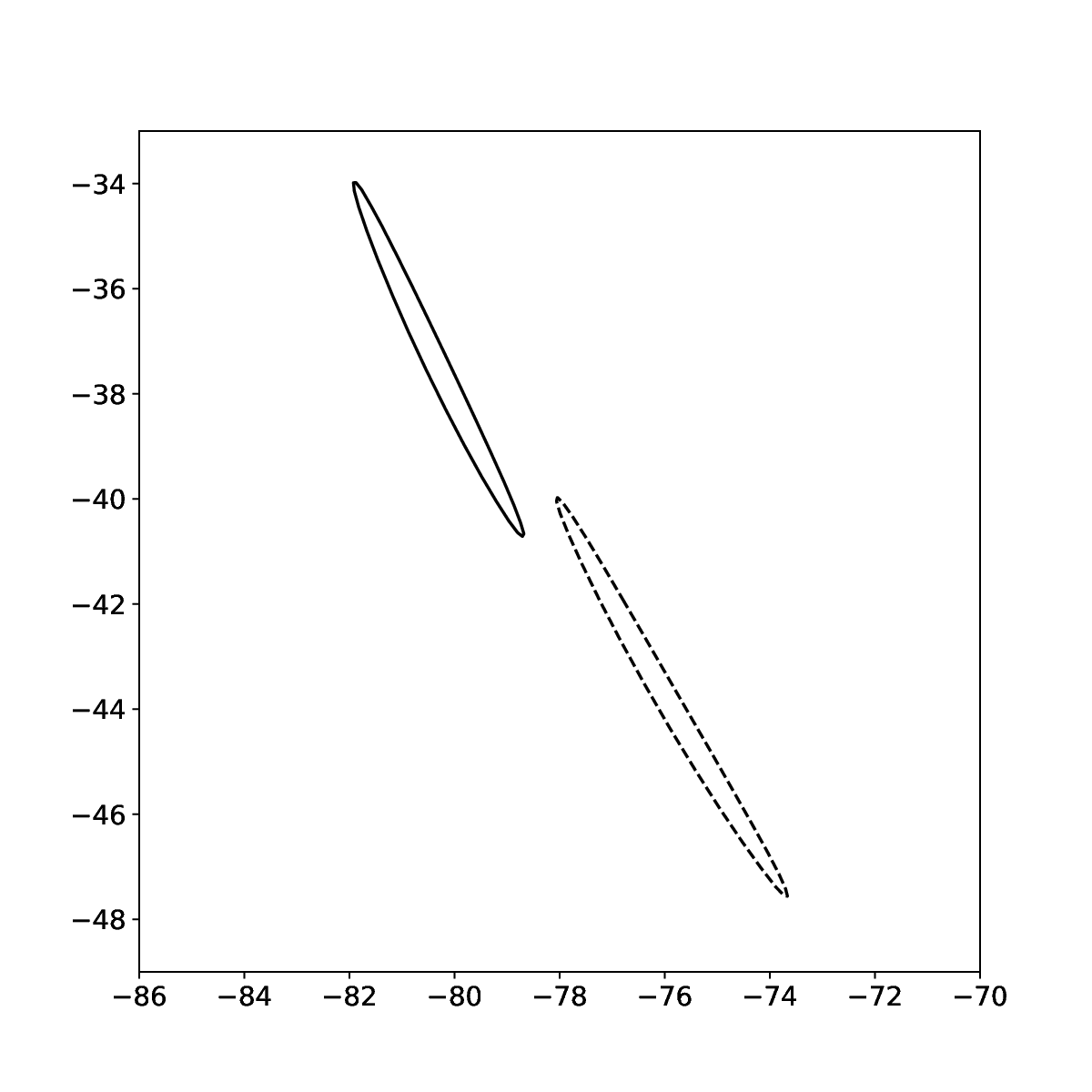}} \caption{\label{extendedarcs:fig}
The gravitational lensed arcs appearing in the lower left quadrant in Fig.~\ref{twoarcs_new:fig} for the source at $\pi/6$. The dashed ellipse corresponds to the $a=0$ case. We see that the rotation of the black hole shifts the position of the image and slightly changes its shape.} \end{center}\end{figure}

In addition to the arcs being shifted in position, the arcs themselves have slightly different appearances in the $a=0$ and $a=0.9$ case. Figure~\ref{extendedarcs:fig} shows the $a=0$ (dashed) and $a=0.9$ (solid) image arc in the lower left quadrant for the source object at $\pi/6$ radians. The presence of rotation shifts the arc position, but also makes the image arc slightly thinner and less elongated.

If we draw a line connecting the centers of the $a=0.9$ image arcs, this line would differ in both slope and intercept from the visualization lines in Fig.~\ref{twoarcs_new:fig}. In the $a=0$ case, the line connecting the two images passes through the black hole center. For images of of the same sources when $a=0.9$, the line connecting the centers of the two images has a different slope and does not pass through the center of the black hole (at the origin). Table~\ref{ArcDiff:table} shows that in the visualization case, that the lines connecting the $a=0.9$ images are substantially different from the line which would connect the images had there been no rotation. The offset in Table~\ref{ArcDiff:table} indicates the distance of closest approach to the origin of the lines connecting the $a=0.9$ images (in the dimensionless units where $M=1$).

At closest approach, the highly elliptical orbit of the star S2 is 120 AU from Sgr A*. The detection of lensed images for source locations near 120 AU behind the black hole may provide a potentially observable test for the effect of rotation. In Table~\ref{SagAoffset:table}, we test lens-source distances of 120 AU and 360 AU, with the observer at our distance from the center of the galaxy. At these scales, the slopes of the lines connecting two image arcs differs from the non-rotating case by less than $1\%$, but the line connecting the two image arc centers is offset from the origin by a distance on the order of micro arc seconds. The distance of closest approach of the connecting line is shown to be sensitive to the value of the spin parameter $a$.

\begin{table}
        \caption{The slopes and distance of closest approach (the offset) of the line  connecting the center of the $a=0.9$ image arcs in Fig.~\ref{twoarcs_new:fig} for the visualization scale. The values for $a=0$ are provided in parentheses. The offset distance is provided in units where $m=1$. }
    \centering
    \begin{tabular}{c|r|r}
    
         position & slope ($a=0$) & offset ($a=0$)   \\
         \hline
         $\pi/6$ & 0.48 (0.58) & 0.79 (0) \\ 
         $3\pi/4$ & -1.35 (-1.00) &  -1.50 (0) \\
        
    \end{tabular}

    \label{ArcDiff:table}
\end{table}

\begin{table}
     \caption{The distance of closest approach to SgrA* of the line connecting the two images of a source located at $\theta = \pi/6$ and $\theta = 3\pi/4$ radians from the horizontal axis when the source is 120 AU or 360 AU behind the black hole and at an axial distances of 20 AU. As the spin parameter increases, the line connecting the two images pulls away from the origin by up to $\sim 3.5 $ micro arc seconds.}
    \label{SagAoffset:table}
    \centering
    \begin{tabular}{c|c|c|c}
         ~ & lens-source &  offset  ($\mu$as)  & offset ($\mu$as) \\
         $a$ & distance (AU) & $\theta = \pi/6 $ &  $\theta = 3 \pi/4$  \\
         \hline
         0.1 & 120 &   0.29 & -0.39 \\
         0.5 & 120 &   1.31 & -1.97 \\
         0.9 & 120 &   2.49 & -3.55 \\
        \hline
         0.1 & 360 &   0.27 & -0.39 \\
         0.5 & 360 &   1.37 & -1.93 \\
         0.9 & 360 &   2.45 & -3.48 \\
    \end{tabular}

\end{table}

The lensed image on the opposite side of black hole from the original source may be significantly de-magnified. For point sources, the magnification factor is determined by the inverse of the determinant of the Jacobian matrix of the lens mapping (\cite{ehlers}). In the context considered here, we have placed the observer in the equatorial plane of the rotation, at $\phi = \pi$, and the traditional lensing ``source plane'' is at a fixed value of $+x = r\sin\theta\cos\phi$. The lens mapping is a mapping from image positions $(y_I, z_I)$, which we project to the source plane location, into true source positions $(y_S, z_S)$, such that the Jacobian matrix is 

\begin{equation} J = \left(\begin{array}{cc} \frac{\partial y_S}{\partial y_I} & \frac{\partial y_S}{\partial z_I} \\ \frac{\partial z_S}{\partial y_I} & \frac{\partial z_S}{\partial z_I} \end{array} \right) . 
\end{equation}

The image on the same side of the lens as the source will be brighter than the image on the opposite side of the lens. The ratio of the image brightness will depend on the ratio of the magnification factors. To estimate the effect, we consider point sources with lens-source distance of 120 AU or 360 AU, located an axial distance of 20 AU and the source, inclined at $\pi/6$ or $3\pi/4$ from the horizontal on the sky, as in Table~\ref{SagAoffset:table}. These flux ratios do not depend strongly on the spin of the black hole, and we report in Table~\ref{fluxRatio:table} flux ratio values for $a=0.5$. We do see a strong dependence on the lens-source distance and some dependence on the position of the source on the sky.

\begin{table}
     \caption{The ratio in brightness of the two images of point sources near Sag A* for the lensing configurations in Table~\ref{SagAoffset:table} for $a=0.5$. Values of the ratio depend strongly on the distance between the lens and source. Ratio values differ by less than $1\%$ between $a=0$ and $a=0.9$. }
    \label{fluxRatio:table}
    \centering
    \begin{tabular}{c|c|c}
         lens-source distance &  angular position  & flux ratio \\
         \hline
         120 AU &   $\pi/6$ & 590 \\
         120 AU &   $3\pi/4$ & 430 \\
        \hline
         360 AU &    $\pi/6$ & 85 \\
         360 AU &  $3\pi/4$ & 76 \\
    \end{tabular}

\end{table}

\begin{figure*}
\begin{center}
\scalebox{0.4}{\includegraphics{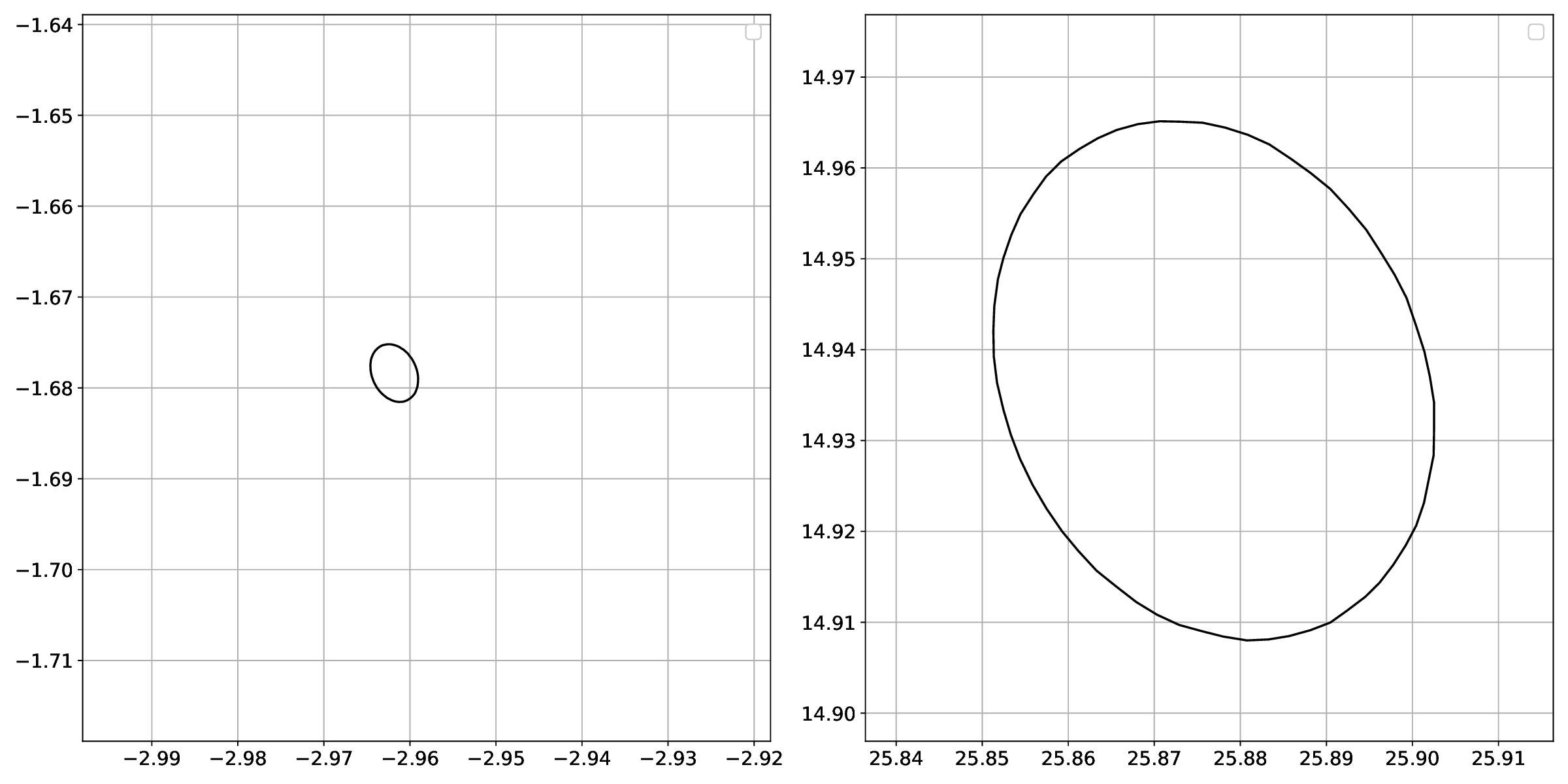}} \caption{\label{lensed360:fig}
The gravitational lensed images for Sgr A* modeled with $a=0.9$ for a lens-source $360$ AU at an axial distance of $20$ AU and $\theta = \pi/6$. The image on the right is the primary image on the same side as the source. The primary gravitationally lensed ellipse is approximately $80$ times larger in area than the image on the opposite side of the black hole. The axes in each plot represent the position relative to Sgr A* in milli arc seconds, and the scale of each plot is the same. } \end{center}\end{figure*}

Figure~\ref{lensed360:fig} shows the two images of a star-sized extended object when the lens-source distance is 360 AU, the axial distance of the source is 20 AU and the source is inclined at $\pi/6$ from the horizontal on the sky. In the right panel, we see the elliptically shaped image of the circular source that is on the same side as the source. The left panel shows the lensed image that has passed by the black hole and appears on the opposite side from the source. The scale of each figure is the same, and we see that the second lensed image is approximately 80 times smaller than the primary image, consistent with Table~\ref{fluxRatio:table}. Because gravitational lensing preserves surface brightness, the significantly smaller second image will be much less bright.

\section{Lensing of Orbiting Objects} \label{orbit:sec}

\cite{doku} have previously shown the gravitational lensing image locations and appearance of a star orbiting very close to a very high spin black hole. The purpose of our work here is to examine how gravitational lensing might impact the primary observable image position under realistic conditions.

\begin{figure}
\begin{center}
\scalebox{0.4}{\includegraphics{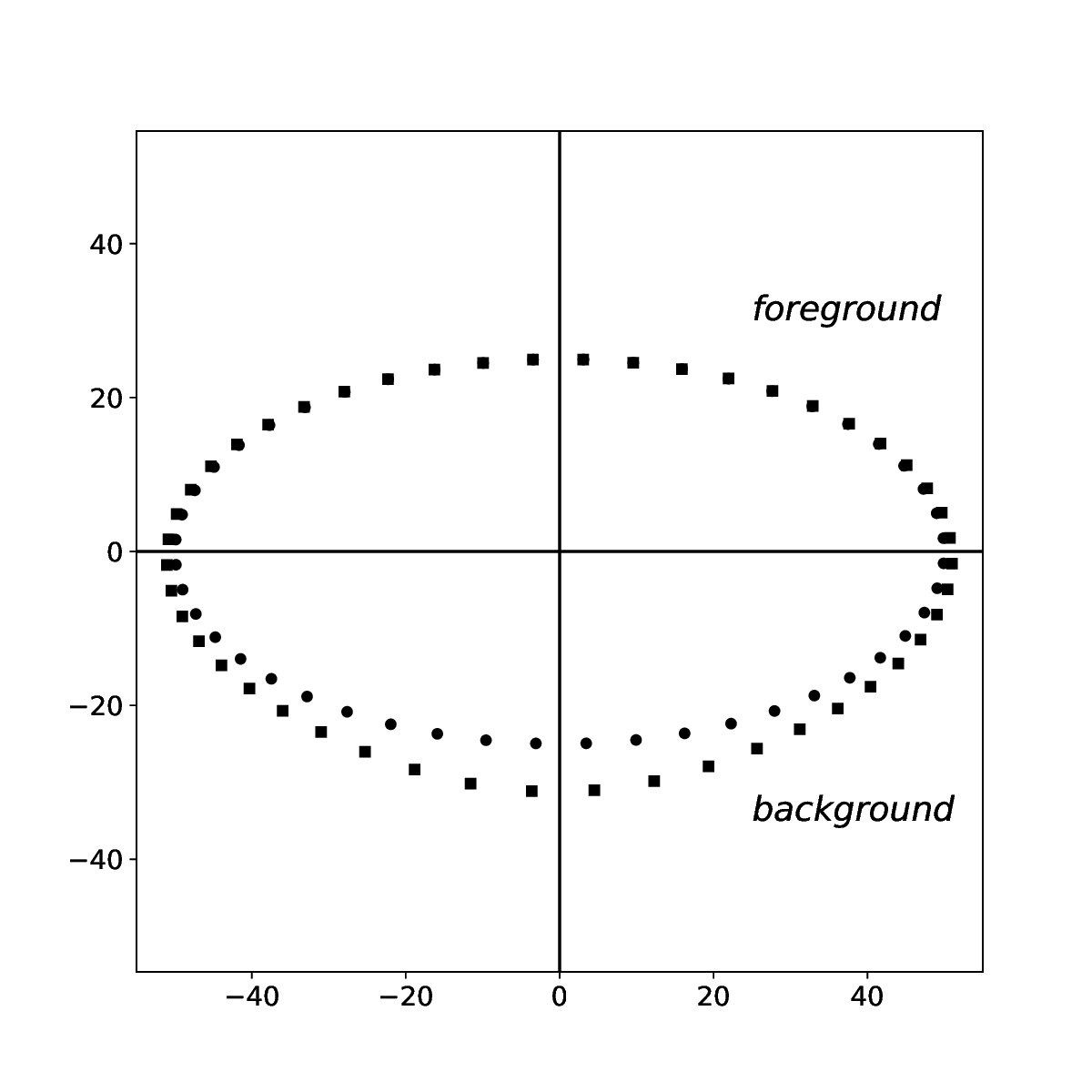}} \caption{\label{orbit1000:fig}
The true sky (circles) and primary image (boxes) locations of an object orbiting a black hole with $a=0.9$ in the visualization scale. The object's orbit is assumed to be circular, with a radius of $50$ scaled units and tipped $30^\circ$ from the vertical such that points in the top half of the figure are in front of the black hole. } \end{center}\end{figure}

To that end, we first visualize the difference between the true source location and the gravitationally lensed image location for an object orbiting a rotating black hole. In Fig.~\ref{orbit1000:fig}, a source is orbiting the black hole in a circular orbit with a radius of $50$ scaled units, tilted $30^\circ$ from the equatorial plane. The observer is at $r_0 = 1000$ in the equatorial plane, the black hole parameters are $m=1$, $a=0.9$. The tilt is such that the source is in the foreground of the black hole when in the upper half of the diagram.

\begin{figure*}
\begin{center}
\scalebox{0.4}{\includegraphics{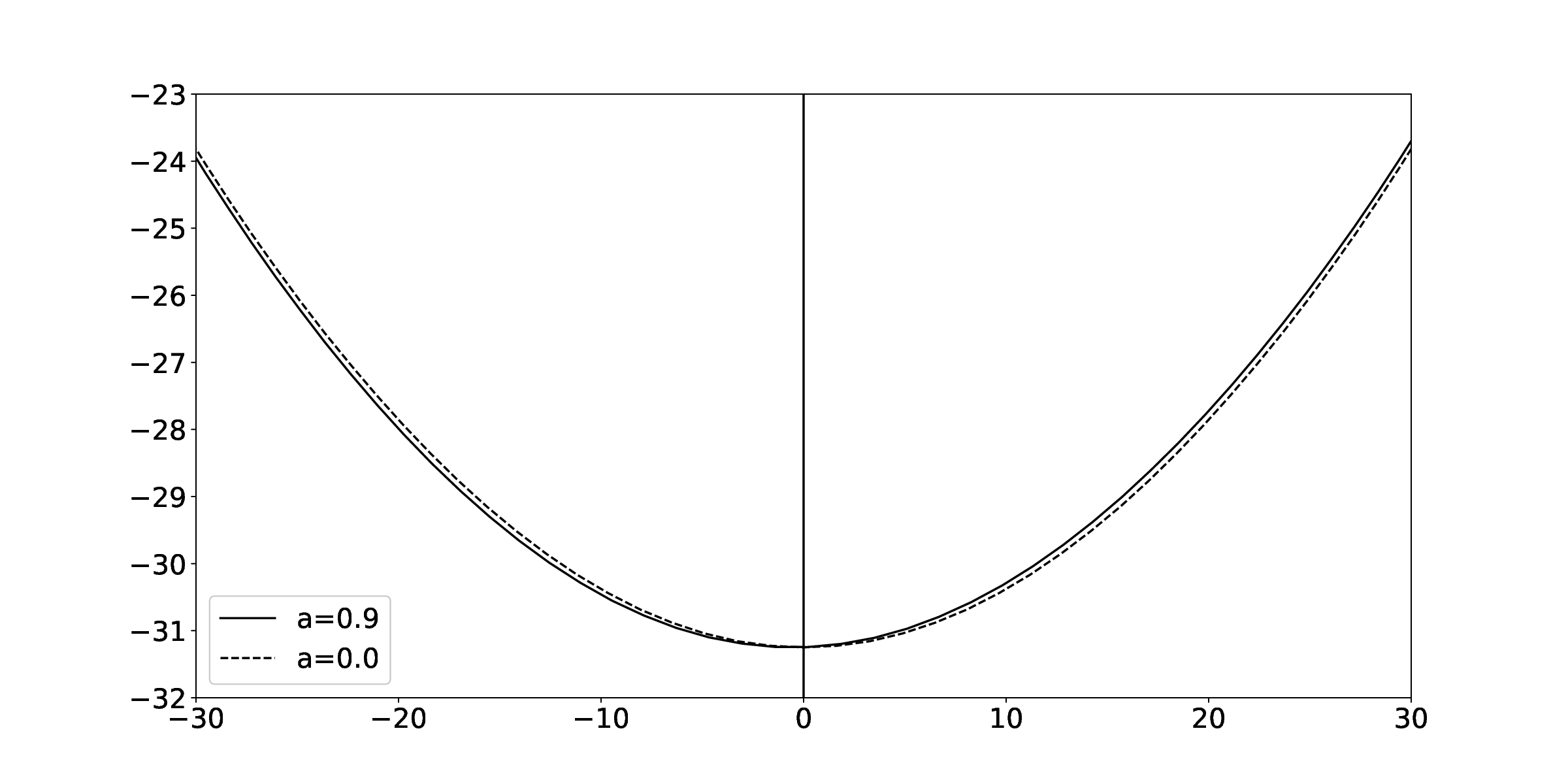}} \caption{\label{orbit1000_witha:fig}
The  primary image locations of an object orbiting a black hole for a portion of the orbit where the source object is behind the black hole. Shown are the image locations with $a=0.9$ (solid curve) and $a=0.0$ (dashed curve) in the visualization scale. } \end{center}\end{figure*}

We see that there are no meaningful differences in the positions for objects in the foreground of the black hole, but that differences in image positions are significant when the source is behind the black hole. Figure~\ref{orbit1000_witha:fig} shows the image locations for the background source positions in Fig.~\ref{orbit1000:fig} but for $a=0$ and $a=0.9$. The black hole spin breaks the symmetry in the image of the orbit.

Table~\ref{S2orbit:table} examines the image locations for an orbiting source in the our true galactic scale. The source positions are chosen to match those in Table~\ref{SagAoffset:table}, with lens-source distances of 120 AU and 360 AU. The stars close to Sgr A* have highly elliptical orbits and interact with each other as well as the central black hole. Our goal is to get a sense of the scale of the shifts between true source locations and gravitationally lensed image locations under some different scenarios. We provide the true source position, the image position (of the brighter image on the same side as the source), and the lensing shift, the distance in angular measure from the image position and the source position.

We see that there are two effects operating on significantly different scales. First, the image location of lensed image is shifted on the order of hundreds $\mu$as for the source positions chosen for essentially all of the $a$ values. We interpret this level of shift as being from the strong lensing due to the black hole mass. Looking at the image positions at different $a$ values, we see an effect that is significantly smaller. Even at very high spin, the shift in the image locations is only about $0.3~\mu$as after taking into account the lensing assuming the black hole has no (or very small) spin. At low spin, say $a=0.1$, the effect is another order of magnitude smaller than at high spin.

\begin{table*}
     \caption{The true source position and observed image positions for different values of $a$ relative to Sgr A*. Source positions are chosen to match those in Table~\ref{SagAoffset:table}. The distance in $\mu$as from the true position and image location is listed as the lensing shift, and the shift between the $a=0$ and non-zero $a$ values increases with the spin parameter.}
    \label{S2orbit:table}
    \centering
    \begin{tabular}{c|c|c|c|c}
         ~ & lens-source &   ~ & ~ & ~\\ 
         $a$ & distance (AU) & true position  ($\mu$as) &  image position  ($\mu$as) & lensing shift ($\mu$as) \\
        \hline
         0.0 & 120 & $\theta = \pi/6 $ & $(2408.17, 1390.37)$ & $115.81$ \\
         0.1 & 120 &  ~ & $(2408.18, 1390.35)$ & $115.81$ \\
         0.5 & 120 &    $(2408.12, 1390.28)$ & $(2587.77, 1494.87)$ & $115.72$ \\
         0.9 & 120 &   ~ & $(2408.10, 1390.20)$ & $115.66$ \\
        \hline
         0.0 &  120 & $\theta = 3\pi/5 $ & $(-1966.28, 1966.28)$ & $115.82$ \\
         0.1 & 120 &  ~  & $(-1966.27, 1966.30)$ & $115.83$ \\
         0.5 & 120 &   $(-1884.38, 1884.38)$ & $(-1966.27, 1966.38)$ & $115.89$ \\
         0.9 & 120 &   ~ & $(-1966.26, 1966.48)$ & $115.98$ \\
         \hline
         0.0 & 360 & $\theta = \pi/6 $ & $(2587.89, 1494.12)$ & $323.33$ \\
         0.1 & 360 &  ~ & $(2587.87, 1494.07)$ & $323.28$ \\
         0.5 & 360 &    $(2307.88, 1332.46)$ & $(2587.77, 1494.87)$ & $323.10$ \\
         0.9 & 360 &   ~ & $(2587.69, 1493.65)$ & $322.92$ \\
        \hline
         0.0 &  360 & $\theta = 3\pi/5 $ & $(-2113.00, 2113.00)$ & $323.33$ \\
         0.1 & 360 &  ~  & $(-2113.00, 2113.05)$ & $323.36$ \\
         0.5 & 360 &   $(-1884.38, 1884.38)$ & $(-2112.97, 2113.30)$ & $323.51$ \\
         0.9 & 360 &   ~ & $(-2112.96, 2113.54)$ & $322.68$ \\
    \end{tabular}

\end{table*}


\section{Discussion}  \label{higher order:sec}

This paper examines the impact of gravitational lensing near rotating black holes, making two unique contributions. First, by visualizing signatures of gravitational lensing with $a\ne 0$, we identify several possible candidates for lensing tests for rotating black holes. Second, we examine the level of accuracy needed in imaging technology to distinguish the gravitational lensing signatures of objects near Sgr A* with and without rotation. 

Figures \ref{SkySource:fig}, \ref{viewShifts:fig}, and \ref{pathsz60:fig} show that for objects sufficiently close to the optical axis connecting the observer and black hole, there can be significant changes in image position. It is not surprising that the rotation breaks the symmetry in image locations, nor that sources that are further away from the optical axis have an image closer to the black hole, and thus test the stronger region of gravity. In fact,  \cite{Bozza05} also noted this effect as a break from symmetry, foreshadowing, but not following up on, the calculation provided in Table \ref{SagAoffset:table}.

Because surface brightness is conserved in gravitational lensing, magnification of images is associated with changes in the size of the image. Figure~\ref{twoarcs_new:fig} shows that when the source is relatively close to the optical axis, the two images are approximately the same size, regardless of $a$ value. As the source location moves further away from the optical axis, for instance the source location marked $B$ in Fig.~\ref{SkySource:fig}, the image that is closer to the black hole does decrease in size. Figure~\ref{lensed360:fig} and Table~\ref{fluxRatio:table} demonstrate that for realistic lensing scenarios, the second lensed image may be hard to distinguish from the background noise as it can be significantly smaller, and less bright, than the main image.

Nevertheless, it is not out of the question that some stars in the background of Sgr A* will cross behind the black hole, close to the optical axis. In these cases, measuring two images should be possible in principle. In this paper, we report that the shape of the images is sensitive to the value of black hole's spin, $a$, and that the line connecting the two images will not pass through the black hole if $a\ne0$. Both of these measurements will be very difficult, requiring seeing in the $\mu$as range with careful examination of signal-to-noise ratios.

We note that the variation in image locations in Table~\ref{S2orbit:table} for different $a$ values is small compared to the overall shift from the true source position. Thus, most of the variation of star position at the $\sim 100\mu$as level is due to the lensing from the mass of the black hole. A high spin ($a=0.9$) black hole affects the image location by about  $0.3\mu$as relative to a $a=0$ black hole. One expects Sgr A* to have a relatively small angular momentum, but future images of sufficient accuracy would be able to distinguish $a$ values. 

However, we also note that the gravitational lensing shift between the true source location and the gravitationally lensed image location is significant and close to the current capabilities with the improved astrometry using the GRAVITY instrument at the ESO VLTI \cite{improved_gravity}. The gravitationally lensing image is, of course, what is observed. When angular resolution reaches this point, dynamical studies that rely on accurate locations of orbiting stars will need to adjust from the observed (lensed) image location to the true source location. Put another way, the observed orbits will have been distorted by gravitational lensing.

The initial modeling of the lensing of stars close to the black hole conducted in this paper should be expanded to take into account full orbit of stars such as S2 to get a full understanding of the impact of lensing.

Ultimately, we conclude that while we are close in the resolution needed to  test the gravitational lensing due to mass of the black hole, another generation of improvements will be needed to measure black hole spin, either through image positions as in Table~\ref{S2orbit:table} or the line connecting images as in Table~\ref{SagAoffset:table}.

\section*{Acknowledgments}

The authors are pleased to acknowledge that the computational work reported on in this paper was performed on the Shared Computing Cluster which is administered by Boston University’s Research Computing Services.  URL: www.bu.edu/tech/support/research/. Lia Troy was supported by the Northeast Cyberteam, through a grant from the National Science Foundation (ACI-1659377). Shannon Harding was supported by a summer research grant from the Massachusetts Space Grant Consortium.

\section*{Data Availability}

Original C++ code using the CUDA library available upon request.



\bibliographystyle{mnras}
\bibliography{references} 


\appendix

\section{Differential Equations for Light Rays}\label{AppA Equations}

The Kerr metric in the $(u_-, r, \theta, \phi_-)$ coordinates is given in Eq.~\ref{kerr2}. Collecting terms in front of each coordinate variation as functions written as $T_i$, we define a Lagrangian for Kerr null geodesics as

\begin{eqnarray} {\mathcal{L}}  & = & \frac{1}{2} g_{ab} \dot x^a \dot x^b  \nonumber \\ &=& T_1 \dot \theta^2 + T_2 \dot r \dot \phi_- + T_3 \dot r \dot u_- + T_4 \dot \phi_-^2 \nonumber \\ & ~ & +  T_5 \dot \phi_- \dot u_- + T_6 \dot u_-^2 , \label{lagrangian} \end{eqnarray}

\noindent  where the dot represents a derivative with respect to an affine parameter $s$. The $T_i$ determined by their presence in front of $dx^a$.

In previous work, it was shown that the resulting Euler-Lagrange equations are coupled, second order ordinary differential equations, but that by algebraic manipulation, the equations could be uncoupled (\cite{klingKerr1}). We introduce several functions to collect terms:

\begin{eqnarray} 
Q_{u_-} &=& -T_{5r} \dot r \dot \phi_- - T_{5\theta} \dot\theta \dot \phi_- - 2 T_{6r} \dot r \dot u_- - 2T_{6\theta} \dot\theta\dot u_- \label{Qu} \\ 
Q_{\phi_-} & = &  -T_{2\theta} \dot r \dot \theta - 2 T_{4r} \dot r\dot \phi_- - 2T_{4\theta} \dot \theta \dot\phi_- \nonumber \\ &~& - T_{5r} \dot r \dot u_- - T_{5\theta} \dot\theta \dot u_- \label{Qphi} \\ 
Q_r & = & -T_{2\theta} \dot\theta \dot \phi_- + T_{1r} \dot\theta^2 + T_{4r} \dot\phi_-^2  \nonumber \\ &~& + T_{5r} \dot \phi_- \dot u_- + T_{6r} \dot u_-^2 \label{Qr} \\
R_1 &=& 2 \left(T_4 + \frac{T_2^2 T_6}{T_3^2} - \frac{T_2 T_5}{T_3} \right)  \label{R1} \\ R_2 &=& Q_{\phi_-}
    - \frac{T_2}{T_3} Q_{u_-} + 2 \frac{T_2 T_6}{T_3^2}  Q_r  - \frac{T_5}{T_3} Q_r \label{R2} 
\end{eqnarray}

\noindent Here, the $r$ and $\theta$ subscripts indicate partial derivatives with respect to those coordinates. 

Because we want to keep track of the physical $\phi$ coordinate which we use for projection into cartesian coordinates, we also integrate the equations for the $t$ and $\phi$ coordinates, but our integration methods applies adaptive stepsize only to the $(u_-, r, \theta, \phi_-)$ equations. Thus, we integrate ten first-order differential equations for Kerr light rays:

\begin{eqnarray} \dot u_-  & = & v_{u_-} \label{udot} \\ \dot r &=& v_r \\ \dot \theta  &=& v_\theta \\ \dot \phi_- &=& v_{\phi_-} \\ \dot v_{u_-} & = & \frac{1}{T_3} \left\{ Q_r - T_2 \left( \frac{R_2}{R_1} \right) \right\} \\
\dot v_r & = & \frac{1}{T_3} \left\{ Q_{u_-} - T_5 \left( \frac{R_2}{R_1} \right) - 2\frac{T_6}{T_3} Q_r \right. \nonumber \\ &~& \left. + 2\frac{T_2 T_6}{T_3} \left( \frac{R_2}{R_1} \right) \right\}\\
\dot v_\theta   & = &   \frac{1}{2T_1} \left( -T_{1\theta} v_\theta^2 - 2 T_{1r} v_r v_\theta +  T_{2\theta} v_r v_{\phi_-}  \right. \nonumber \\ & & ~~ \left. + T_{4\theta} v_{\phi_-}^2 + T_{5\theta} v_{\phi_-} v_{u_-} + T_{6\theta} v_{u_-}^2 \right) \\
\dot v_{\phi_-} & = & \frac{R_2}{R_1} \label{phiminusdot}
\\ \dot t & = & v_{u_-} + \frac{(r^2 + a^2)}{\Delta} v_r \label{tdot} \\ \dot \phi &= &v_{\phi_-} + \frac{a}{\Delta} v_r \label{phidot}. \end{eqnarray}

For initial conditions, setting an angle $\alpha$ between initial direction of the light ray and the line connecting the observer and origin allows one to set the values of $v_{\phi_-}$ and $v_\theta$ as initial directions of the light ray that can be freely chosen. Thus, these two parameters span the sky of null directions at the observer.


\section{Integration Techniques for Black Hole Lensing}\label{AppB Integration}

Our approach to gravitational lensing for Kerr black holes is to set the observer location and source position and integrate light rays as ordinary differential equations as described above. The ethos of the exact lensing approach is to consider past light rays of the observer, and that the lens map is a map of the sky of the observer onto a distant source plane where individual light sources reside. For us, the initial conditions at the observer have two free parameters, $v_\theta$ and $v_{\phi_-}$, that parameterize the sky of the observer. When a source location is set, the goal is to location the set of $\{ v_\theta, v_{\phi_-}\}$ values that create light rays connecting observer and source. Because we are doing black hole lensing, there are an infinite number of pairs of such values, as the light rays can circle the black hole many times. Because of the black hole spin, $a$, pairs of $\{ v_\theta, v_{\phi_-}\}$ values are not symmetric, and each image must be found individually. 

To simplify the expression of the lens equation, we introduce standard cartesian coordinates $(x,y,z)$, and for this paper, we consider the observer to be at a position on the $x$ axis at $-x_0$, for $(\theta = \pi/2, \phi = \pi)$. The sources are considered to lie in a source plane defined by a constant, usually positive $x = x_s$ value. The lens mapping is then a mapping from the $\{ v_\theta, v_{\phi_-}\}$ space into the $x_s$ plane where sources are at $(y_s, z_s)$. Image locations are created by projecting the $\{ v_\theta, v_{\phi_-}\}$ into the $x_s$ plane assuming no lensing ($m=0, a=0$).

A given $\{ v_\theta, v_{\phi_-}\}$ pair defines a light ray, and assuming that the source plane at $x_s$ is relatively far from the black hole, say $x_s > 5$, the light ray will intersect the source plane no more than one time. (If the light ray falls into the black hole, or if it turns around the black hole and escapes in a direct not generally along the line of sight, it will not intersect the source plane at all. We did not examine sources close enough to the event horizon that a single light ray intersects the source plane more than one time.) For light rays that do intersect the source plane, we can define a simple distance function $d = \sqrt{(y_s - y_\ell)^2 + (z_s - z_\ell)^2}$ between intersection point of the light ray $(y_\ell, z_\ell)$ and a specified source location $(y_s, z_s)$. Finding $\{ v_\theta, v_{\phi_-}\}$ values connecting the observer and source is then the problem of finding the zeros of the function $d$.

In their famous book \textit{Numerical Recipes}, \cite{NRC} comment that there are no good methods for finding the roots of such functions. This is the case for Kerr black hole lensing. The nonlinearity of the differential equations means that desired roots lie at a set of global $d=0$ minima troughs, where the steepness of troughs is very uneven along different directions. We found that following the ``downhill'' direction by taking the local gradient of the distance function numerically was ineffective due to the quick changing nature of $d$ near roots for light rays that pass near the black hole. Creating a refining a ``square'' grid of $\{ v_\theta, v_{\phi_-}\}$ proved more effective but slow.

Our method for solving the lens equation involves three steps. First, we create a forward mapping by integrating a very large number of light rays at varying $( v_\theta, v_{\phi_-})$ from the observer until the rays reach the source plane, $x_s$. In the visualization scale, we typically integrate $40,000$ light rays, but when modeling Sgr A*, or when we examine light ray paths closer the the black hole, we increase the size of the forward map to up to $360,000$ rays. Because we use NVIDIA Tesla V100 gpus for this calculation, increasing the forward map size does not significantly impact performance time directly. For each ray in the forward map, we record the initial sky position, $(v_\theta, v_{\phi_-})$, and final source position $(y_\ell, z_\ell)$. 

With the forward map in place, as a second step we compute the distance to the source location for each element in the forward map, and then sort map by smallest distance ($d$) values. (At this step, an increased size of the forward map does slow the process.) Within the sorted forward map, we select up to six $(v_\theta, v_{\phi_-})$ pairs that have the smallest value of $d$ taking account that we want \textit{distinct} initial directions and that these will not simply be the six smallest values of $d$ within the forward map. The selected initial conditions connecting the source to locations near the source point are a backward map. 

Finally, we refine the backward map to find $(v_\theta, v_{\phi_-})$ pairs that reach the source location within a specified tolerance.  The tolerance used in this paper was $0.00316$ dimensionless units (where $m=1$). As a result, in the visualization scale our error from refining light rays is $0.3$ arc seconds. Because the tolerance is kept the same for the Sgr A* simulations, light rays connect the observer and final source position to within $0.02~\mu$as.

Our backward map refinement process also involves integrating light rays from the observer to the source plane, but this time in bundles near the initial values. At the center of the bundle is the light ray determined from above to connect the observer to a point close to the desired light source. We use 12 points in an elliptical pattern in $(v_\theta, v_{\phi_-})$ space to trace the outline of the bundle. Each light ray is integrated to the source $x$ distance, and distances to the desired source are computed. If a point on the edge of the bundle is the closest to the desired $(y_s, z_s)$, the center of the bundle is moved to this $(v_\theta, v_{\phi_-})$ and the ellipse of the next bundle is made larger a factor of three. If the original point in the center of the bundle is the closest to the desired $(y_s, z_s)$, the ellipse size is shrunk by a factor of one third. We stop the process either after the distance $d$ is less than a proscribed tolerance (success) or attempting to move the bundle a large number of times (failure). Through this process, the bundle slides over and finds the minimum. We find that in practice, convergence to our desired tolerance is achieved after 30 to 50 recursions if the initial positions are reasonably close, and we are able to resolve light rays connecting the observer and source that wrap around the black hole multiple times.

\section{Verification and Validation}\label{AppB Integration}

Because the results of this paper are generated by numerical integration of light rays, it was important to establish that the numerical modeling was accurate. We first verified that the integration of the geodesics in Eqs.~\ref{udot}-\ref{phidot} with $a=0$ matches the integration of geodesics of a Schwarzschild metric, independently coded from the Schwarzschild metric in $(t, r, \theta, \phi)$ coordinates. Then, we wrote the Lagrangian function of the Kerr metric in Boyer-Lindquist coordinates into Mathematica, and used Mathematica's symbolic algebra capabilities to create the second-order Euler-Lagrange equations. The output of the C++ integration of our equations from this paper were then compared with Mathematica's integration (using NDSolve) of the Euler-Lagrange Equations generated through Mathematica's symbolic calculus procedures. We verified that our numerical integration of our equations, in $(u_-, r, \theta, \phi_-)$ coordinates, agreed with Mathematica's integration of its generated equations for light rays in Boyer-Lindquist coordinates, providing a strong test of both the numerical integration techniques we used as well as our basic equation structures.  

\begin{figure}
\begin{center}
\scalebox{0.4}{\includegraphics{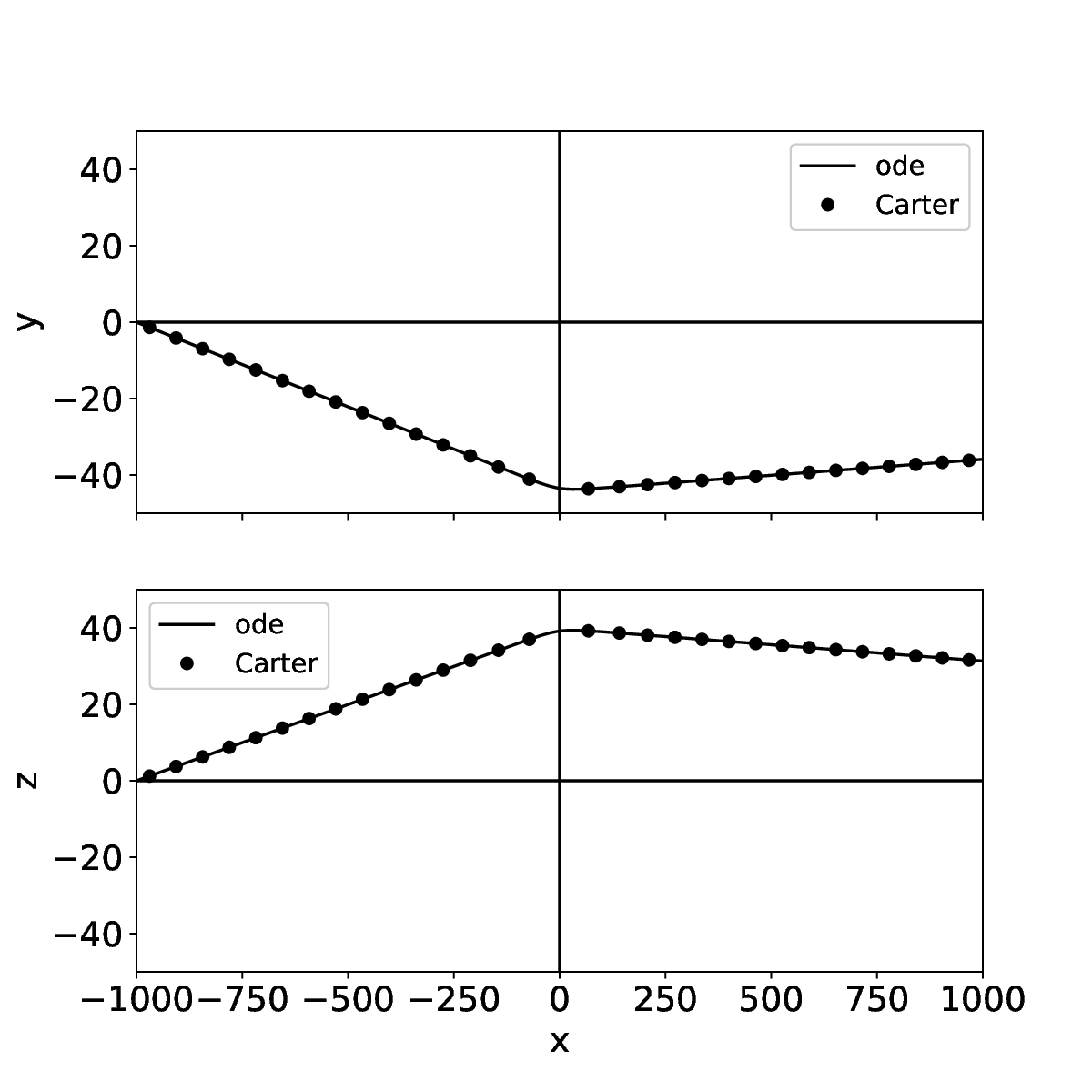}} \caption{\label{Carter:fig}
Projections into the $x-y$ and $x-z$ plane of the path of a light ray in an $m=1$, $a=0.9$ Kerr black hole spacetime as computed using the ODE approach of this paper (solid line) and the common integral approach (solid dots). } \end{center}\end{figure}

Finally, we were interested in comparing our approach to integrating the light rays of the Kerr black hole space-time with the more common integral approach introduced by \cite{carter} and used by many authors. In particular, we followed the approach outlined in \cite{Gralla1}. To do so, one must carefully align the initial conditions (of our ODEs) and the constants of integration in the integral approach. Figure~\ref{Carter:fig} shows the spatial path of a light ray in the visualization scale with $m=1$ and $a=0.9$. With appropriately matched initial conditions and constants of integration, the two methods match.

\bsp	
\label{lastpage}
\end{document}